\RequirePackage{fix-cm}
\documentclass[smallextended]{svjour3} 

\smartqed

\usepackage{graphicx}

\usepackage[cmex10]{amsmath}

\usepackage{parskip}

\usepackage{fixltx2e}

\usepackage{upgreek}

\usepackage[authoryear]{natbib}

\usepackage{aas_macros}

\usepackage[hyphens]{url}

\date{} 
\journalname{Experimental Astronomy}

\begin{document}

\title{Performance Analysis of GPU-Accelerated Filter-Based Source Finding for HI Spectral Line Image Data}
\titlerunning{Performance Analysis of GPU-Accelerated Filter-Based Source Finding}
\author{Stefan Westerlund \and Christopher Harris}
\institute{Stefan Westerlund \at
              ICRAR/The University of Western Australia, M468 35 Stirling Highway, Crawley, WA 6009 \\
              Tel.: +61 8 6488 1780\\
              \email{stefan.westerlund@icrar.org}           
           \and
           Christopher Harris \at
              iVEC@UWA/The University of Western Australia, M024 35 Stirling Highway, Crawley, WA, 6009\\
              \email{christopher.harris@uwa.edu.au}
}

\date{Received: date / Accepted: date}

\maketitle

\begin{abstract}
Searching for sources of electromagnetic emission in spectral-line radio astronomy interferometric data is a computationally intensive process. Parallel programming techniques and High Performance Computing hardware may be used to improve the computational performance of a source finding program. However, it is desirable to further reduce the processing time of source finding in order to decrease the computational resources required for the task. GPU acceleration is a method that may achieve significant increases in performance for some source finding algorithms, particularly for filtering image data. This work considers the application of GPU acceleration to the task of source finding and the techniques used to achieve the best performance, such as memory management. We also examine the changes in performance, where the algorithms that were GPU accelerated achieved a speedup of around $3.2$ times the 12 core per node CPU-only performance, while the program as a whole experienced a speedup of $2.0$ times.


\end{abstract}

\section{Introduction}

Future radio astronomy surveys will produce large amounts of data, in the petabyte scale and beyond. This data must be processed and analysed in order to identify the galaxies, and other sources of emission, that are present and to measure their properties. Doing so using consumer desktops or laptops is infeasible due to the computational, memory, and storage requirements of the data. High Performance Computing (HPC) resources are necessary to process this scale of data in a reasonable amount of time.

This work concerns the use of source finders, which search a radio astronomy image for the galaxies and other objects it contains. The source finding techniques presented in this work are designed for searching spectral image cubes that are produced by blind surveys of HI emission, and cover a large spatial and spectral extent. There are source finders for this task that already make use of HPC resources, but it is beneficial to further increase the processing speed of these programs. Reducing the computational resources needed for source finding allows a survey to reduce its computational costs, use them for other tasks, or to use more accurate but more computationally intensive algorithms.

Previous work on this subject, the Parallel Gaussian Source Finder (PGSF) \citep{hpc_source_finding} searches images by applying multiple three-dimensional Gaussian filters to the image data. These filters are used to reduce the effects of noise on the image, making the signal of the emission sources more clear. A local calculation of the mean and standard deviation of the filtered data is used to create thresholds for locating sources. PGSF employs the Scalable Source Finding Framework (SSoFF) \citep{hpc_source_finding} for parallelism, which in turn uses the Message Passing Interface (MPI) library \citep{mpi} to communicate information between different processes on different computing nodes. The distribution of work across multiple processes allows the source finder to make use of additional processors and memory and thus make use of cluster and supercomputing resources. However SSoFF can only make use of CPU architecture.


Graphics Processing Units (GPUs) have been shown suitable for other computationally intensive steps in the radio astronomy spectral-line image processing pipeline. Signal correlation  has been the target of GPU acceleration efforts, including \citet{gpu_ocl_signal_correlation} and \citet{gpu_acc_signal_convolution}, and GPU acceleration is being used in the software correlator for the Murchison Widefield Array \citep{mwa_software_correlator}. GPUs are also used for de-dispersion of fast transient radio astronomy signals \citep{realtime_gpu_transient_dedispersion}, including software detection pipelines developed for the Advanced Radio Transient Event Monitor and Identification System survey (ARTEMIS) \citep{artemis_gpu_fast_transient} and the Commensal Realtime ASKAP Fast Transient Survey (CRAFT) survey \citep{craft_gpu_dedispersion}. Previous work has also shown that GPUs are capable of accelerating signal and image processing algorithms, including data clustering \citep{gpu_clustering} and edge detection \citep{canny_edge_detection}. Because these works show that tasks that have similar computational requirements to source finding can be successfully accelerated using GPU computing, it is likely that source finding is also a good target for GPU acceleration.

This work examines the use of GPU acceleration, applied to the task of source finding. The most time-consuming portions of the code were identified and ported to run on GPUs, using the OpenCL and CUDA frameworks. This work considers a range of GPU computing techniques employed to gain optimum performance from the hardware, such as memory management and data transfer. The performance of the two GPU implementations was measured along with the performance of the original CPU implementation, so that the speed of the different versions could be compared to each other. Other technical issues encountered in creating the GPU accelerated program and measuring its performance are also discussed.

The following section describes the background to the existing MPI-based program, identifies the target portions for acceleration, and introduces the relevant GPU computing terminology. Subsequently, the Method section describes the manner in which the source finding tasks were ported to the GPU, including data distribution and optimisation techniques, along with modifications to algorithms that were not ported to GPU. The Results section then reports the performance of the source finder under different conditions, including the performance the accelerated filter algorithm, the overall speed of the source finder and changes in the portion of processing time for different algorithms on the program. The Discussion section analyses the results and compares the different GPU and CPU implementations. Finally, the Conclusion section summarises the results.

\section{Background}
In our previous work \citep{hpc_source_finding} we created the Parallel Gaussian Source Finder (PGSF), which is accelerated using MPI to make use of multiple CPUs. In this section we first consider the performance of this existing program to determine the components that are suitable for GPU acceleration. We then provide a background to the mechanics of GPU acceleration in preparation for the subsequent sections.


\subsection{Existing Sourcefinder Framework}
The PGSF program described in \citet{hpc_source_finding} uses a series of three-dimensional Gaussian filters to locate sources. To improve performance PGSF uses MPI to communicate across multiple processes, allowing the program to take advantage of multiple computing nodes. Using additional nodes increases the number of CPUs available to process the data, and increases the amount of memory that is available for PGSF to store information.

The source finding task is parallelised by splitting the image cube up into chunks, and giving each chunk to a different MPI process. That process is then responsible for analysing that portion of the image. This includes communicating its results to the processes that are analysing neighbouring data and combining the results to a single, parameterised catalogue at the end of the search. PGSF takes advantage of parallel supercomputing hardware to achieve good performance, but it is preferable to further increase the performance of the program in order to reduce the amount of time and computing resources required to search an image.


In order to speed up a program its most time-consuming portions must be identified, so that optimisation efforts can be concentrated there. This is because speeding up the slowest part of the program will have the greatest effect on the overall run time of the program. Preliminary profiling of PGSF shows that the most time consuming portion of the program is the filtering process. This task applies several filters to the data using convolution. If $d$ is the image data, $w$ is the weights data, $x$, $y$, and $z$ are indices into the images, $f$ is the filter with a radius of $r_{x}$, $r_{y}$, and $r_{z}$ then the convolved image, $c_{w}$ can be calculated as shown in Equation~\ref{weighted_conv_eqn}. If a weights image is unavailable then an unweighted convolution can be performed, calculating $c_u$ according to Equation~\ref{conv_eqn}.

\begin{align}
	\label{weighted_conv_eqn}
	c_{w}[x][y][z] &= \frac{\sum\limits_{o_x} \sum\limits_{o_y} \sum\limits_{o_z} d[i_x][i_y][i_z] ~ f[o_x][o_y][o_z] ~ w[i_x][i_y][i_z]}{\sum\limits_{o_x} \sum\limits_{o_y} \sum\limits_{o_z} w[i_x][i_y][i_z]} \\
	\label{conv_eqn}
	c_u[x][y][z] &= \sum\limits_{o_x} \sum\limits_{o_y} \sum\limits_{o_z} d[i_x][i_y][i_z]~ f[o_x][o_y][o_z] \\
	i_x &= x + o_x, ~o_x \in [-\frac{F_x}{2},\frac{F_x}{2}] \\
	i_y &= y + o_y, ~o_y \in [-\frac{F_y}{2},\frac{F_y}{2}] \\
	i_z &= z + o_z, ~o_z \in [-\frac{F_z}{2},\frac{F_z}{2}]
\end{align}

Given the range of radio astronomy and signal and image processing techniques utilising GPU-accelerated programming discussed in the previous section, this filtering algorithm is a good candidate to be improved using GPU computing. Before presenting our method, we first provide a brief introduction to the relevant GPU concepts.

\subsection{GPU Acceleration}
To accelerate the filtering algorithm, two GPU frameworks were investigated in this work. These are CUDA \citep{CUDA} and OpenCL \citep{opencl}. Compute Uniform Device Architecture (CUDA) is a framework made by NVIDIA and runs only on their GPUs. Open Computing Language (OpenCL) is a framework created by the Khronos group. OpenCL can run on a variety of devices with implementations available for CPUs, GPUs, and other types of processors. As CUDA and OpenCL use different terms to refer to the same concepts, this paper will use CUDA terminology.

Both CUDA and OpenCL are examples of what is known as \emph{heterogeneous computing}. This is where different types of devices are used together to calculate a result, taking advantage of the advantages of each type of device. In the case of using GPUs to assist general computing, the program runs on the CPU, which is called the \emph{host} and sends data and commands to the GPU, which is called the \emph{device} and performs the accelerated computing.

The executable code that runs on the device is called the \emph{kernel}. Each thread on the GPU runs an instance of the kernel. Threads are grouped together into \emph{blocks}, which can work together. In the case of NVIDIA GPUs, the processor is divided into a number of \emph{streaming multiprocessors}, or SMs. Each SM consists of a number of \emph{cores} that execute a single thread of the kernel. At any given time, all of the threads on a SM run in lock step, executing the same part of the same kernel. This group of threads is called a \emph{warp}, the size of the warp is equal to the number of cores in a SM and is important for performance considerations. All of the threads in a warp must be part of the same block, and have successive block indices. The collection of blocks that run a kernel is called a \emph{grid}. Blocks and grids may have indexing for up to three dimensions. Threads in the same block can communicate with each other, including using barriers to control program flow. Threads in different blocks cannot communicate with each other, even if they are in the same grid.

Memory is divided up into several different locations. The memory of the host device is separate from any memory on the device. The \emph{global memory} is the main memory bank of the device, and all threads can access it, albeit with a latency of hundreds of clock cycles, depending on the hardware \citep{cuda_programming_guide_5_5}. \emph{Shared memory} is a memory space that is shared between different threads in the same block, and is significantly faster than global memory with accesses times on the order of tens of cycles. \emph{Local memory} is memory space that can only be seen by a single thread, and is also fast. Both shared and local memory are located in the same symmetric multiprocessor as their threads. Some devices provide a small space for \emph{constant memory}, which can be used to efficiently broadcast data to large numbers of threads.

There are several factors that can affect the speed of memory transfers between global memory, and either shared or local memory. If the threads in a block are accessing successive values in global memory, then it is possible that the transfer can use \emph{coalesced reads and writes}, where multiple values are sent or received in a single transaction. The particular requirements and benefits of coalesced reads and writes vary greatly depending on the hardware and software versions used. If multiple threads are accessing the same values in memory then the kernel may encounter \emph{bank conflicts}, where the threads must wait for previous memory transfers to complete, as they use the same communication channels.

Even when using coalesced reads and writes and writes, intra-device communication can still be much slower than computation, so devices employ a technique called \emph{latency hiding}. This is where multiple threads can wait to read or write data whilst other threads are executing on the cores. The ability of the device to perform latency hiding is limited by the \emph{occupancy} of the kernel, which is how many threads can be kept in a SM's memory at the same time. The occupancy is determined by the resources of the symmetric multiprocessor and the resource requirements of the kernel. All of these features must be considered when writing efficient GPU code. The techniques used to apply GPU acceleration to the filtering function of  the source finder are explained in the next section.




\section{Method}
In order to improve the speed of a program, the most time-consuming components of the algorithm must be identified. Previous work on PGSF shows that the filtering section is the most computationally intensive stage of the source finder, taking up as much as $70\%$ of the total running time. Other significant time-consuming tasks are the statistics calculations and reading in the image and weights data. The file input time is dependent on the storage system and network hardware, and is not significantly affected by changes in algorithm. As the filtering is the most time-consuming task of the source finder, it is the main focus of optimisation efforts. The techniques used to accelerate the filtering algorithm are explained below, including data partitioning and management, and the implementation of the filtering kernel. The statistics calculations are also considered in Section~\ref{stats_parallel}, followed by technical considerations in Section~\ref{section_technical_considerations}.

It is worth noting that while the CPU-only version of PGSF uses one single-threaded MPI process per core, the GPU-enabled implementations of PGSF use one MPI process per node. This is done for ease of programming when managing the GPU resources. In cases where more CPU processing power is needed by the program, OpenMP is used to execute code on the remaining CPU cores on each node. This aspect is expanded on in Section~\ref{stats_parallel}.


\subsection{Data Partitioning and Transfer}

The first issue encountered when porting an algorithm to GPUs is memory management. The host and device have separate memory spaces so any data to be processed must be sent to the GPU, and the results of the computation need to be copied back to the host. This is complicated by the fact that the two different platforms have different hardware capabilities, in particular the GPU has significantly less memory than the host machine. Just as a single node cannot contain an entire image in memory, a GPU cannot hold the entire amount of data held by a single node. Therefore the image data, as well as any weights data, will need to be divided up into smaller sections that can be sent to the GPU and processed one at a time.



The convolution algorithm used to apply the filter to a particular element of the image data requires the image and weights information of all the other voxels within a range equal to the size of the filter in each of the three dimensions of the data. This means that the image can be divided up into three-dimensional sections of consecutive values such that all the data needed for one voxel can be found elsewhere in the section. There will need to be some overlap, or \emph{halo data} between these sections so that the GPU can filter the voxels that are at the edge of the section. These sections can be sent to the GPU one at a time, filtered and then the results returned to the host before the next section is processed.

\begin{figure}
	\centering
	\includegraphics[width=\columnwidth]{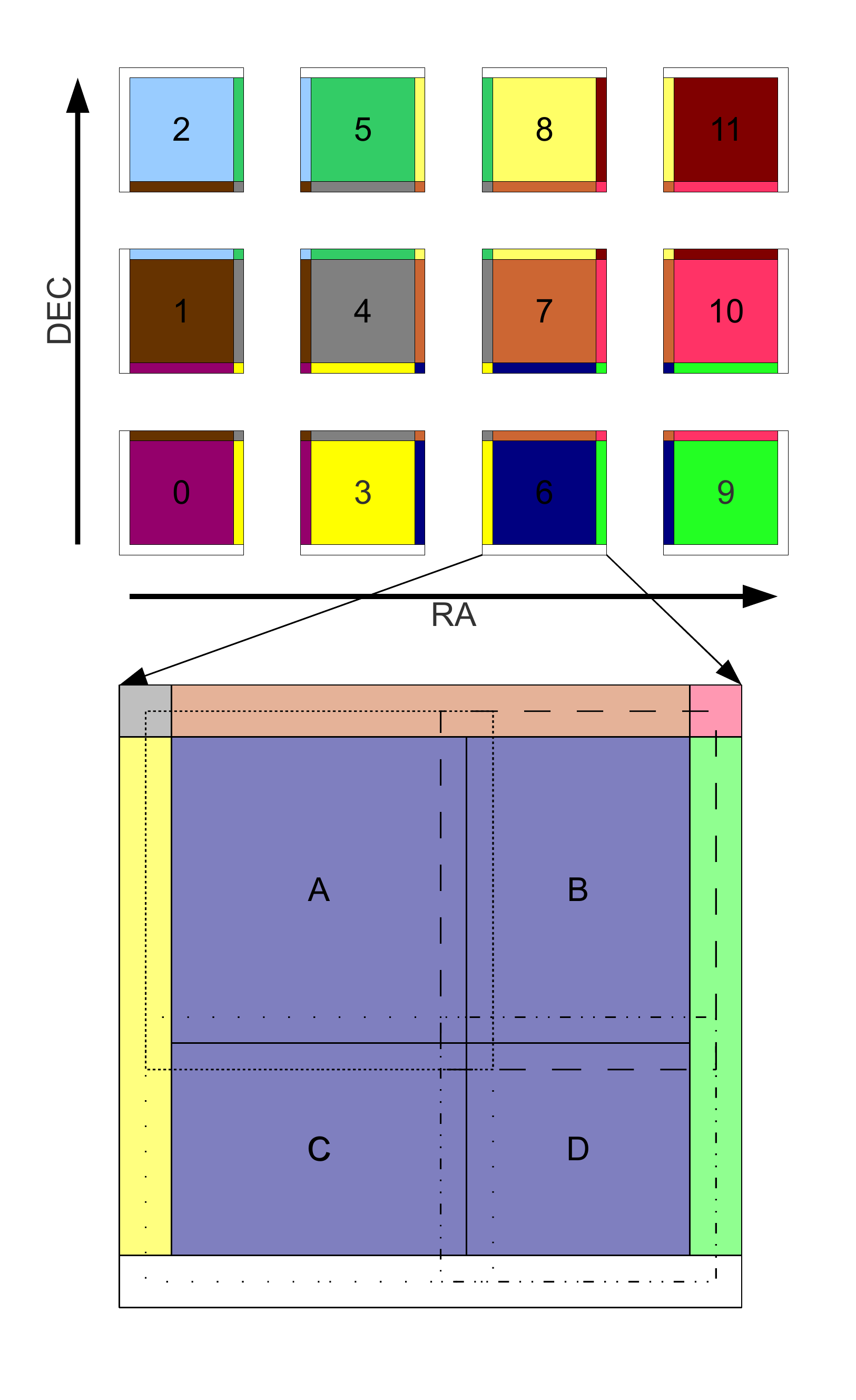}
	\caption{{\bf Data Distribution.} This figure shows the data of a node being divided up into four sections: A, B, C and D. Each section has a portion of data that will be filtered, shown by the solid lines for each section, and a halo of data that is needed for filtering the section but not filtered itself. The halo data is shown in the fine dotted line for section A, the dashed line for section B, the coarse dotted line for section C, and the dot-dashed line for section D. The manner in which the node's data is split into sections is unrelated to the division of the image data between nodes. Instead, the program chooses the sections to be as large as possible given the amount of memory available on the device, until the end of the data is reached.}
	\label{data_distribution}
\end{figure}

Dividing the data into sections for the device adds an extra layer of data distribution underneath what is already present in PGSF. When distributing the image data between different nodes each node is given an equal-sized portion of the image to process, to achieve an even load distribution. The node stores the data for their portion of the image, along with some halo data from surrounding nodes, as needed. When dividing a node's data into sections that can fit onto the GPU, the size of these sections is chosen to minimise the overhead from the data transfer. This further division of image data used by the program to send data to the GPU is shown in \figurename~\ref{data_distribution}. This figure only shows the X and Y axes for clarity. The host data is also divided up along the Z axis according to a user-specified parameter, while for the GPU division only entire lines of spectral data are sent to the device. Both the host and device memory in the  order of X as the major axis and Z as the minor axis.



The maximum size of the sections is limited by the amount of memory available on the device. The device needs to be able to hold one section that contains the unfiltered values, a second section that stores the results of the filter, and when weights are being employed a third section is needed to store the filter weights. Each of these sections contains an inner core which is the set of voxels that are being processed and is variable in size. The section also contain an outer halo of values that is equal to the size of the filter, and needed by the GPU to process the voxels at the edge of the core of the section. Although the section used to store the filtered output does not need to store any halo data, it is kept at the same size as the input and weights sections so that all three sections can be addressed using the same indexing. The device also needs to store the filter values, program code and other constants, but these are insignificant compared to the size of the sections.

Larger sections result in a smaller amount of halo data relative to the total size of the image. This reduces the total amount of data being sent to the GPU and therefore the amount of time spent transferring data. For this reason it is desirable to use the largest section size possible, although the improvement has diminishing returns as the section size increases. The host program checks the amount of memory available on the device being used and combines that with the number of sections required to determine the maximum number of voxels that can be stored at once. From there, the program determines the maximum possible size of the sections, including the halo data needed. The exception to this is when the edge of the node's data is reached. In this case, only the remainder of the data is sent to the GPU in a section, as shown in the smaller sections on the right and the bottom in \figurename~\ref{data_distribution}. The reason for this is that re-arranging the data to send a larger section to the GPU would require more time than what would be gained by using a large section.

There are some complications to the size of the section that comes from the frameworks being used. OpenCL devices have a limit for the maximum size of a single buffer, and this limit may be significantly less than the total amount of global memory available on that device. It is possible that this limit may constrain the maximum size of a section used by the source finder, particularly in the case of an unweighted data set. This limit does not exists on the CUDA platform. However, CUDA provides a function that allocates memory for a two- or three-dimensional array in memory using extra values to pad its contents to meet the alignment requirements of the device. The extra memory used in padding the array must be considered when calculating the maximum possible size of the sections.



Each portion of the image data is processed one section at a time. Sending the data to the GPU, applying the filter, and reading the results back all occur sequentially. It is possible to arrange the algorithm such that the data transfers overlap with data processing on the GPU, and to overlap transfers to the device with transfers from the device. However, testing shows that the data transfer time is approximately $5-10\%$ of the time spent filtering the data. This is true for both the CUDA and OpenCL implementations of the filtering process, so using overlapped communication would not significantly improve the processing speed of the source finder.

The image, weights and output data sent to and read from the GPU is stored in the GPU's global memory. Other values are stored in the GPU's constant memory, including the filter values and the sizes of the arrays being used. The manner in which these values are transferred to the processing cores of the GPU, and how the results are calculated from the data, is discussed in the following subsection.



\subsection{Filter Parallelisation and Implementation}
The filtering is the component of the program that is optimised using heterogeneous computing techniques, by porting the filtering algorithm to run on GPUs. As seen in Equations \ref{weighted_conv_eqn} and \ref{conv_eqn}, each element of the output image can be calculated independently of the other elements in the convolved image using a gather operation, so each element in the image can be processed independently. This results in a maximum parallelisation of one thread per element in the image, and so it can benefit from an increasing number of cores. The organisation of these cores in processing the data greatly affects the efficiency of the kernel.

\begin{figure}
\centering
	\includegraphics[width=\columnwidth]{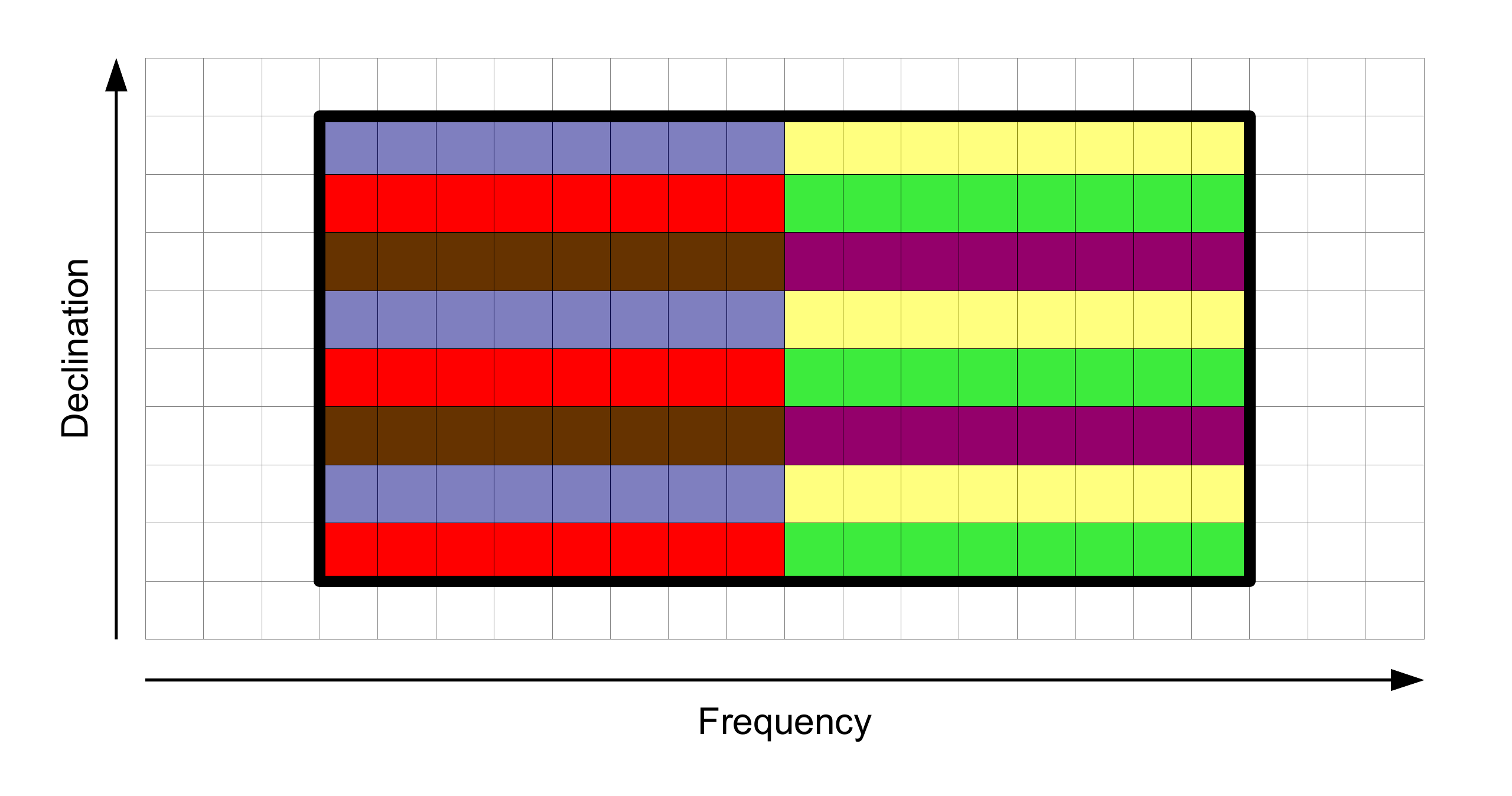}
	\caption{{\bf GPU Thread Assignment.} The grid shows the image data stored in the GPU's global memory. The section surrounded with the heavy black line is the portion of data that will have its filtered value calculated in this kernel. The colour shows the different blocks assigned to process that portion of the data. Here each block is $1 \times 1 \times 8$ threads in size. For clarity, this figure only shows the declination and frequency axes of the data.}
	\label{gpu_processing_distribution}
\end{figure}

The sources are longest in the spectral dimension, relative to the resolution of the images, and correspondingly the filters are also largest along the spectral dimension. Therefore the kernels will achieve the greatest benefit from being aligned along the spectral axis, with each thread responsible for calculating the filtered output for consecutive frequency voxels. The threads are arranged in blocks that are $1 \times 1 \times N$ threads along the x, y and z dimensions of the data, which correspond to the right ascension, declination and frequency axes of the image. The distribution of processing among the different threads and blocks is shown in \figurename~\ref{gpu_processing_distribution}. The parameter $N$ can be controlled by the user, and the effect it has on performance is considered below.

The blocks can be used to read an entire line of frequency values into shared memory, which is a number of values equal to the number of threads plus the halo values needed at the start and end of the line. The image data, and weights data if any, is ordered such that successive values in memory are different frequency values along the same spectrum so the block can read in a line of frequency values using coalesced reads. Employing shared memory in this fashion means that the data from each coalesced read can be used a number of times equal to the number of filter elements along the frequency axis. It is possible to arrange the blocks so that they use shared memory to store a greater portion of the image at once, but this causes some problems with performance. Increasing the amount of memory required by a block will reduce the number of blocks that can be held in memory at once, which reduces thread occupancy and the ability of the GPU to perform latency hiding. Second, if the amount of shared memory used by a block is too large it will spill over into global memory, which removes any performance advantage of using shared memory. In order to achieve the best performance, the amount of memory stored by each block is limited.

The size of the block has several properties that affect the performance of the kernel. As the size of the block increases, the greater amount of shared memory available allows it to reduce overhead from reading in the halo data of the image and weights sections. However, the increased memory use also decreases the maximum occupancy of the blocks, and so reduce the amount of latency hiding the GPU can perform. Additionally, it is preferable that the total number of threads in a block is a multiple of the symmetric multiprocessor size, so that the blocks fit evenly into the hardware. The optimum block size is dependent not only on the variables listed here and the kernel being used, but also the version of the CUDA library and the particular hardware being used.

\begin{figure}
	\centering
	\includegraphics[width=\columnwidth]{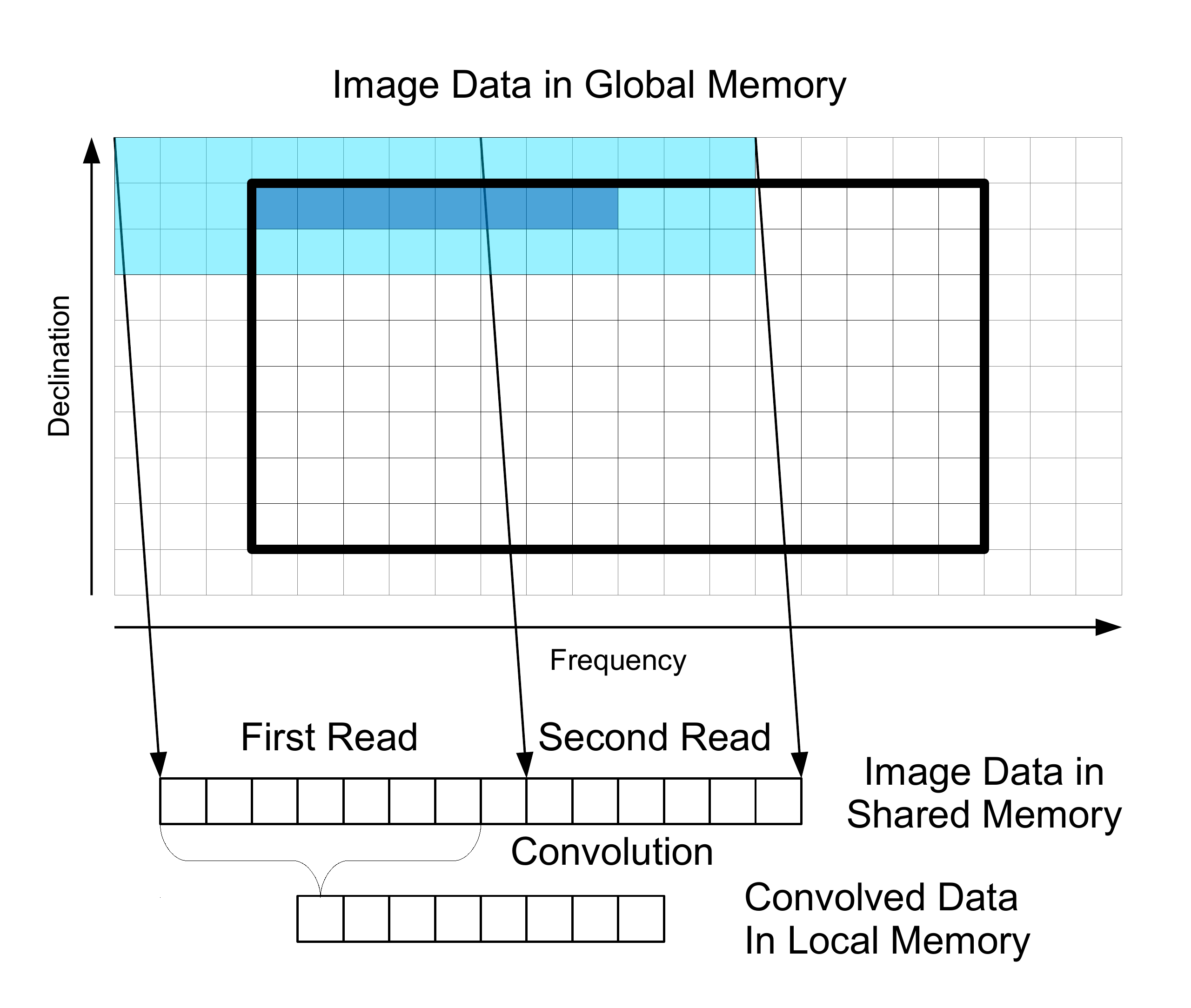}
	\caption{{\bf Kernel Memory Access.} Each block reads the image data, and any weights data, from global memory into shared memory one spectrum at a time. Once the data has been loaded into shared memory each thread uses those values to calculate the convolved sum for the loaded portion of the image and adds it to that voxel's total, which is stored in local memory. The dark blue section is the set of voxels whose filtered values are being calculated by this block of threads. The light blue shows the region of data that is needed to calculate those filtered values.}
	\label{gpu_memory_pattern}
\end{figure}

Using this block topology, it is possible to write the kernel that calculates the filter convolution for the data. The same algorithm and optimisations are used for both the CUDA and OpenCL implementations of the filtering. Each thread starts by calculating its index in the output segment, which determines the particular voxel that thread will be filtering, and initialises its convolution sum. Then for each X and Y position in the filter, the block reads in the line of frequency values that it needs from the image data. The particular line of data read is based on the X and Y position for which the threads are calculating the output, plus the X and Y offset for the filter element that the block is currently calculating. Because the block needs to read in a larger number of image values than the size of the block, it will take two or more reads for the block to acquire all of the values it needs. The memory read patterns used by the kernel are shown in \figurename~\ref{gpu_memory_pattern}. If a weighted convolution is being performed, then a line of weights values are also loaded into the block's shared memory.

The data read is followed by a barrier to ensure that all of the data has been loaded before it is used by any of the threads. Once the barrier has been passed, each thread performs the filtering process for that line. Each image value, and the relevant weights value, is read from shared memory and the appropriate filter value is read from constant memory. The values are multiplied together and added to the thread's convolution sum. Note that at each time step the threads in each warp are reading an offset image and weight value, but read the same filter value. Once a thread has processed all the values along the line, it reaches a second barrier, to prevent the next loading stage from overwriting the block's shared memory before the other threads in the block have finished with the line. After the barrier the threads continue the loop for the next X and Y line in the filter. When the calculation is complete the results are written to the output section. As the threads in the block refer to consecutive elements in the array, this is done using coalesced writes.

Whilst the data in global memory and shared memory is stored as single-precision floating point values, the convolution sums and the convolution calculations are carried out using double-precision floating point values. This sacrifices some speed for greater accuracy in the filter calculations. The difference in processing speed will vary significantly with the GPU hardware and libraries used, in addition to the kernel implementation. There is minimal difference in the memory requirements for using double precision calculations, as only one or two variables are in double precision. Depending on the accuracy requirements of the survey data being examined by PGSF, it may be possible to switch to single-precision calculations to speed up the filtering without a significant decrease in the final accuracy of the detections.

It is possible to arrange the blocks to process only one output voxel per kernel, or to have them filter multiple output voxels. Assigning a block to process a smaller number of threads increases the total number of threads queued to run on the GPU, increasing the maximum possible parallelism. Increasing the total number of threads may also increase the scheduling overhead, decreasing performance. Like the size of the block, the performance effects of changing the number of voxels each block processes may also vary according to the version of CUDA and the hardware being used. The number of voxels each block processes may be controlled by the user, so that it can be adapted for optimum performance.

\subsection{Statistics Parallelisation}
\label{stats_parallel}
The mean and standard deviation values used by PGSF to determine the detection threshold are calculated independently for each voxel from a user-specified range of image data around the voxel being calculated, after the image data has been filtered. This choice is significantly more computationally expensive than calculating a single mean and standard deviation for the entire dataset. The implementation of the local statistics calculation used by PGSF was modified to employ OpenMP to allow the program to make use all the cores in a node. PGSF used one MPI process per CPU core, so all of the cores were used even though only a single thread per core. The GPU implementations only use a single MPI process per node, so the use of OpenMP allows the program to regain use of all of the CPU cores on the node. This increases the performance of the single MPI process per node versions of the program up to the speed of the CPU-only, one MPI process per core implementation, and does so without the effort of porting the local statistics function to OpenCL and CUDA. OpenMP is used to parallelise the outer loop of each of the three directional sums, copying data to temporary buffers, and calculating the z-score values.

\subsection{Technical Considerations}
\label{section_technical_considerations}
Some devices employ a feature called a \emph{watchdog timer}. This will stop a compute kernel if it is still executing a set time after it started, under the assumption that the kernel has crashed or is in an infinite loop. This is problematic because by default the entire set of image data on the device is executed in a single kernel, thus it is possible that the watchdog timer will terminate the filtering kernel. In some cases the device can be configured to disable the watchdog timer, but this is not always possible, particularly if the source finder is being run on a shared supercomputing system. This program works around the watchdog timer by splitting the filtering work across multiple kernel executions, such that each kernel runs in less than the watchdog time. The maximum amount of processing to perform per kernel execution can be set by the user.

Constant memory is used to store the filter data and the sizes of the arrays. This allows the data to be broadcast to multiple threads, when those threads were accessing the same memory location, avoiding the bank conflicts that would occur if global memory was used.

\section{Results}

This program was tested using Fornax, a 96-node GPU-accelerated cluster run by iVEC. Each node has two six-core 2.66GHz Xeon X5650 CPUs, one NVIDIA Tesla C2075 GPU and 72GB of memory. The nodes are connected by two QDR Infiniband networks, one for MPI traffic and the other for storage traffic. The storage system uses the Lustre file system with $32$ storage nodes.

This program is designed to be capable of processing data produced by the Australian Square Kilometre Array Pathfinder (ASKAP) telescope \citep{askap}, in particular the Widefield ASKAP L-band Legacy All-sky Blind surveY (WALLABY) \citep{wallaby_paper}. This survey has yet to start, so there is not yet any real data to test. Instead a simulated image is used, that is $2048 \times 2048 \times 4096$ voxels 64GB in size. It has been concatenated four times to match the $2048 \times 2048 \times 16384$ voxel size that the actual ASKAP images are expected to be.

The tests of the source finder were performed in two parts. First the filter kernel parameters were optimised in order to find the best-performing settings. Then the performance of the overall program, including the optimised kernels, was measured as the number of nodes used to run the source finder was varied over consecutive tests.

\subsection{Kernel Testing}
\begin{figure}
	\centering
	\resizebox{\columnwidth}{!}{\input{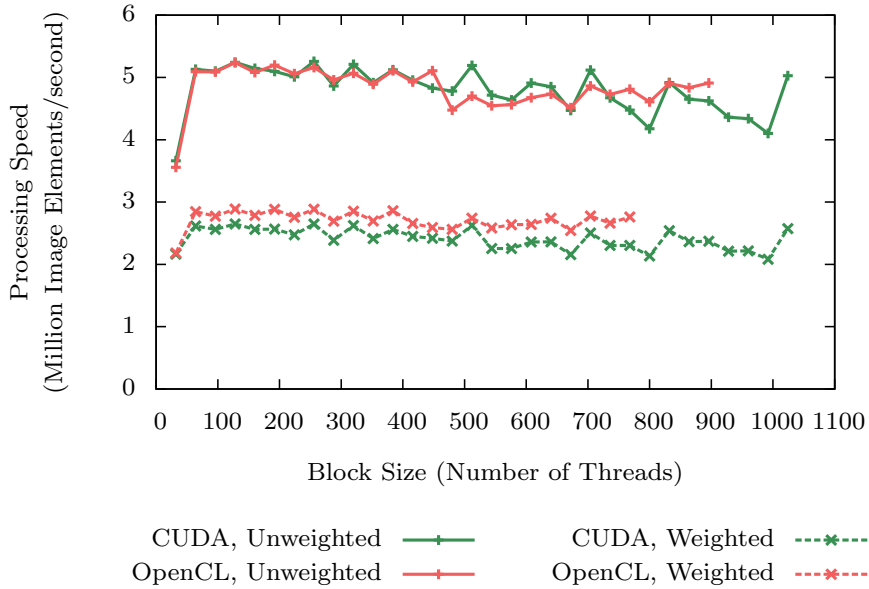}}
	\caption{{\bf Block Size.} This shows the processing speed of the program when using different block sizes in OpenCL and different block sizes in CUDA. The processing speed is a relative measure. The maximum block size is limited by the amount of memory needed by each thread.}
	\label{group_size}
\end{figure}

The performance of the kernels was measured across several different block sizes, in order to determine the optimum value for the hardware. In order to prevent unnecessary consumption of supercomputer time, these were calculated using a cutout of the simulated image that is $1024 \times 1048 \times 4096$ elements in size, so it was small enough to fit onto a single node. The block sizes tested are multiples of 32, which is the warp size for the GPUs used and the preferred block size multiple as reported by OpenCL. Three tests were run per parameter, the mean of the filtering time of each was used as the result and shown in \figurename~\ref{group_size}. The best results for OpenCL were obtained using a block size of $128$ for both the unweighted and weighted convolution. For CUDA the optimum values were block size of $128$ for unweighted and $256$ for weighted convolution. A number of other block sizes achieved almost equal performance, particularly around block sizes of $128$, $256$, $320$, and $512$. The optimum value for each category was used in the subsequent tests.

\begin{figure}
	\centering
	\resizebox{\columnwidth}{!}{\input{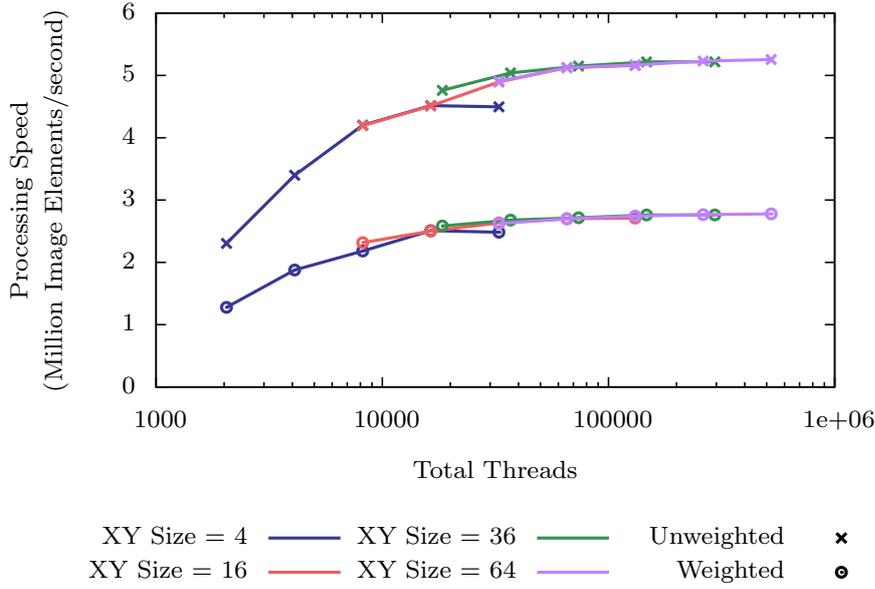}}
	\caption{{\bf OpenCL Total Threads.} This plot shows the relative processing speed of the filtering as a function of the total number of threads. The total number of threads can be specified in three dimensions, with different thread counts in the x and y directions being shown in different colours, and different z counts in the same colour.}
	\label{ocl_total_threads}
\end{figure}

\begin{figure}
	\centering
	\resizebox{\columnwidth}{!}{\input{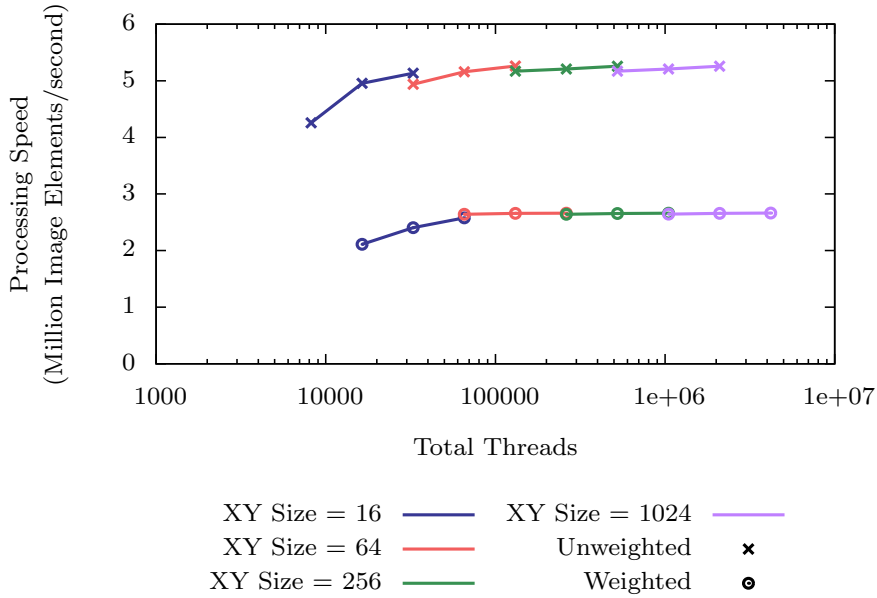}}
	\caption{{\bf CUDA Total Threads.} This plot shows the relative processing speed of the CUDA filtering function as a function of the total number of threads. The total number of threads can be specified in three dimensions, with different thread counts in the x and y directions being shown in different colours, and different z counts in the same colour.}
	\label{CUDA_total_threads}
\end{figure}

The effect of the global thread count on the performance of the program was also measured. Like the block size, these tests were performed using a single node and a cutout image, with three tests per parameter. The performance as a function of the global thread count is shown in \figurename~\ref{ocl_total_threads} for the OpenCL implementation and \figurename~\ref{CUDA_total_threads} for the CUDA implementation. The X and Y size of the blocks is the number of threads in the X direction, which maps to right ascension in the image, multiplied by the number of threads in the block in the Y direction, which maps to the declination of the image.

The tests with CUDA were analysed using the CUDA profiling tools. In particular, they revealed that the data transfer to and from the device took less than $10\%$ of the computation time when applying the filter and that the thread occupancy was $80-83\%$ of the theoretical maximum. Additionally, the profiler shows that the kernel is compute-bound, rather than IO-bound so improving data transfers on the kernel is unlikely to significantly improve processing time.

\subsection{Program Testing}
\begin{figure}
	\centering
	\resizebox{\columnwidth}{!}{\input{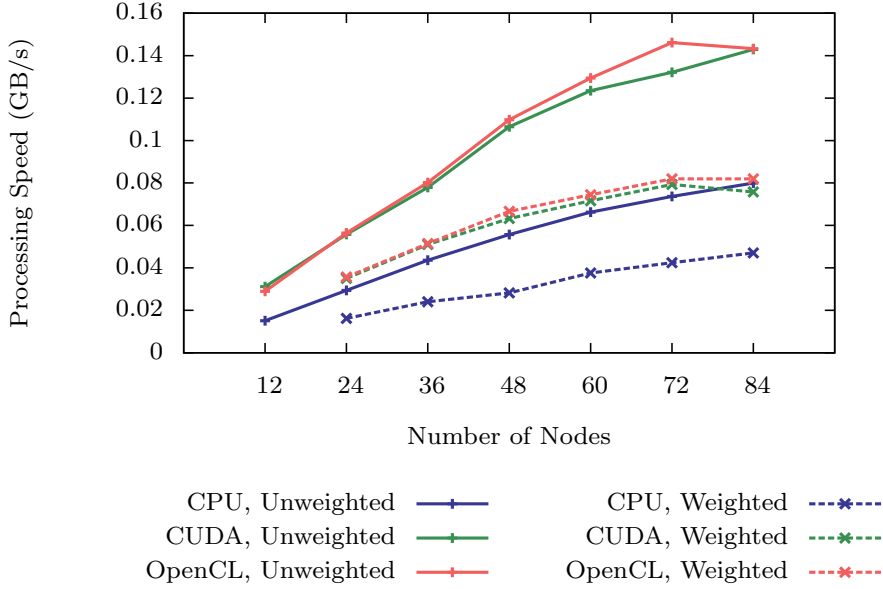}}
	\caption{{\bf Process Scaling.} This plot shows the computational performance of the source finder as a function of the number of computing nodes used in the processing. This plot shows different of the filtering algorithm, with both weighted and unweighted convolutions and implementations using OpenCL, CUDA or the CPU. The OpenCL and CUDA implementations use one CPU core per node in addition to the GPU, the CPU implementation used twelve core per node throughout. The dataset used is a 256GB simulated image.}
	\label{process_scaling}
\end{figure}

The scalability of the program was measured by running the source finder on increasing numbers of nodes, as shown in \figurename~\ref{process_scaling}. These tests perform ten runs for each implementation of the program and node size, and the mean of each set of runs is used. This data measures the performance of the entire source finding program, not just the filtering. The weighted implementations of the program were unable to complete when using only twelve nodes due to memory constraints. The file input time used for each combination of implementation and number of nodes is the maximum recorded input time for all of the runs in that combination. This makes the input time highly susceptible to outliers, but it provides the closest value to what can reasonably be expected when the input time can vary greatly between runs. In addition the time of each step was measured independently, as in \citep{hpc_source_finding}, so they can be examined individually.

\begin{figure}
	\centering
	\resizebox{\columnwidth}{!}{\input{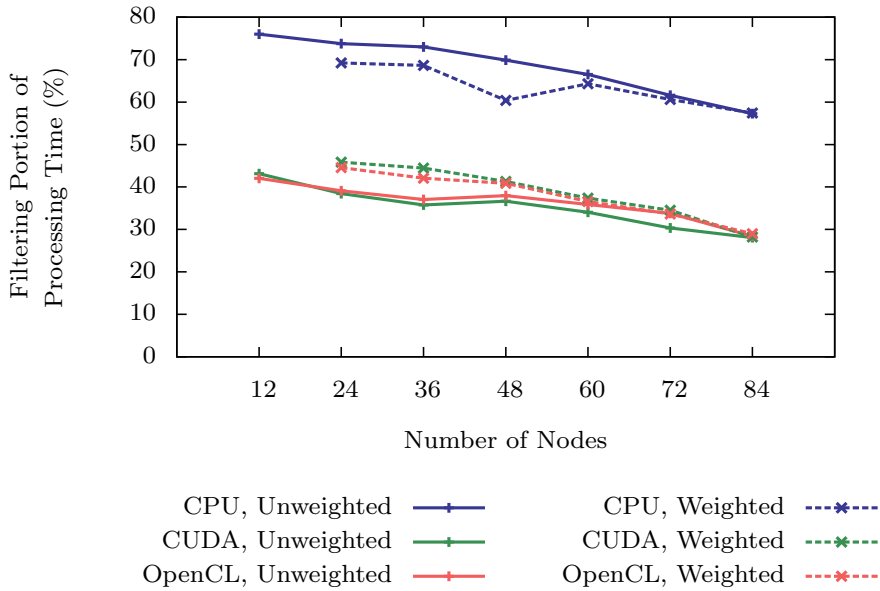}}
	\caption{{\bf Filtering Time Portion.} This plot shows the portion of time each implementation spent applying the three-dimensional Gaussian filters to the image and weights data, as a percentage of the overall run time.}
	\label{filtering_time_portion}
\end{figure}

\begin{figure}
	\centering
	\resizebox{\columnwidth}{!}{\input{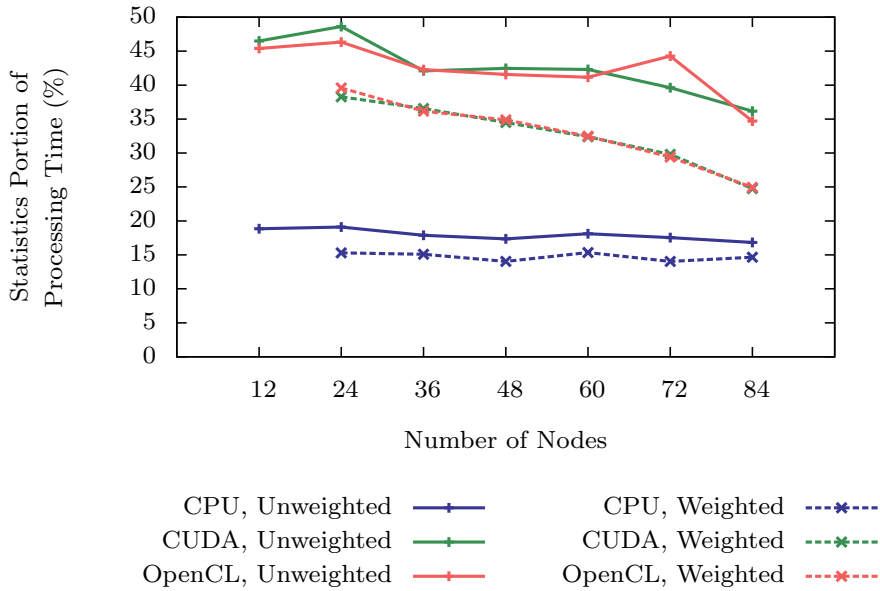}}
	\caption{{\bf Statistics Time Portion.} This plot shows the portion of time each implementation spent calculating the local mean and standard deviation of the filtered image data, as a percentage of the overall run time.}
	\label{statistics_time_portion}
\end{figure}

\begin{figure}
	\centering
	\resizebox{\columnwidth}{!}{\input{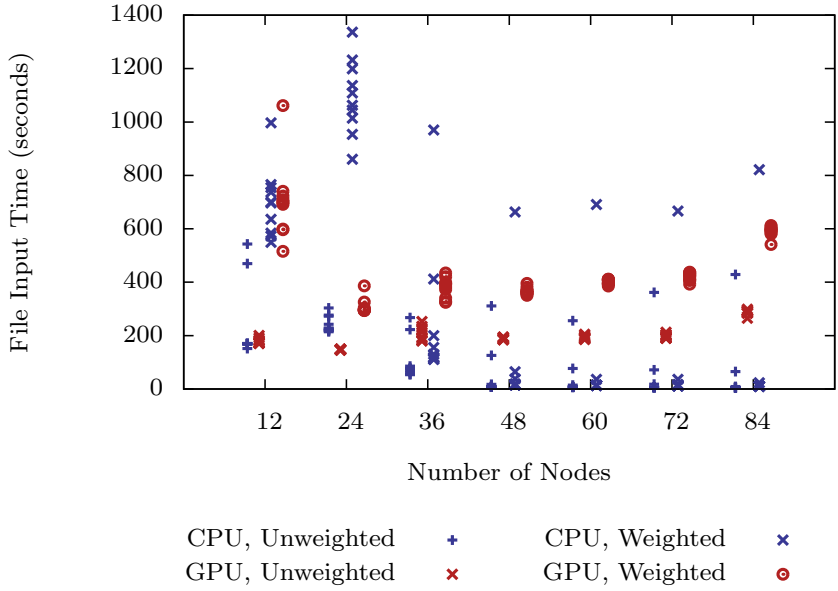}}
	\caption{{\bf Input Time.} This plot shows the time taken by the source finder to read in the image data. For the unweighted case, this is just the 256GB image file. For the weighted case, this includes reading in both the image file and an additional 256GB weights file. In both cases the time also includes the time taken to communicate the halo data between different processes.}
	\label{input_time}
\end{figure}

\begin{figure}
	\centering
	\resizebox{\columnwidth}{!}{\input{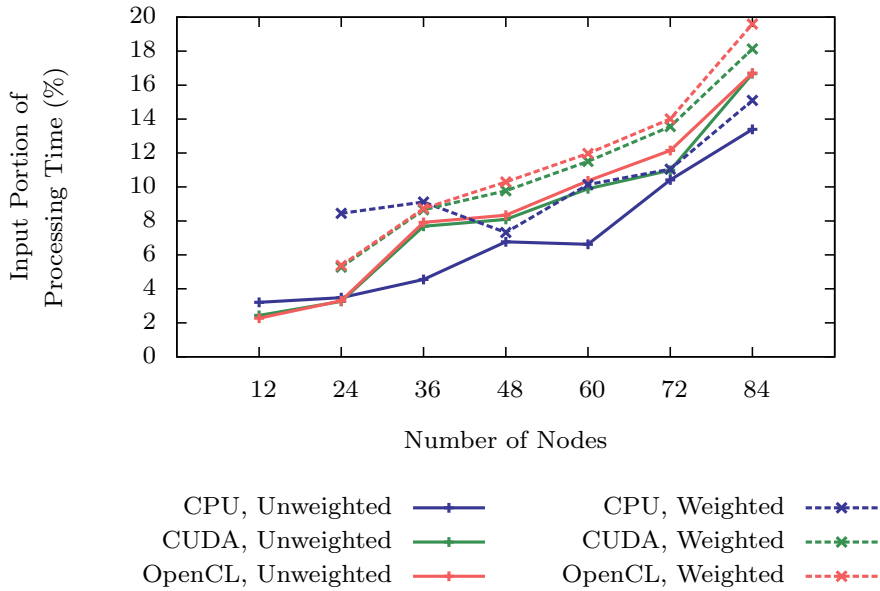}}
	\caption{{\bf Input Time Portion.} This plot shows the portion of time each implementation spent reading in image and weights data, as a percentage of the overall run time. This figure uses the maximum input time of each run for that combination of node count and implementation.}
	\label{input_time_portion}
\end{figure}

Because the input time varies so greatly, the scatter of input times was also measured. The times taken to read the input data is shown in \figurename~\ref{input_time}. For the unweighted implementations this measures the time taken to read in the image file, for the weighted implementations this records the time taken to read both the image and weights files. By comparing the maximum input time against the overall time taken to search the image, the extent to which the input time comprises the overall processing time can be seen. This information is shown in \figurename~\ref{input_time_portion}.

During testing of the program a drop in IO performance was encountered, which was identified to be due to a change in configuration in the Lustre shared file system. Once this issue was corrected the input performance was tested again. In order to reduce the amount of computer resources spent testing the program, only the input stage of the program was run. One set of tests was run for the CPU implementation, and a second set of tests was run for the GPU implementations. This can be done as the input step for the two GPU implementations is identical.


\begin{figure}
	\centering
	\resizebox{\columnwidth}{!}{\input{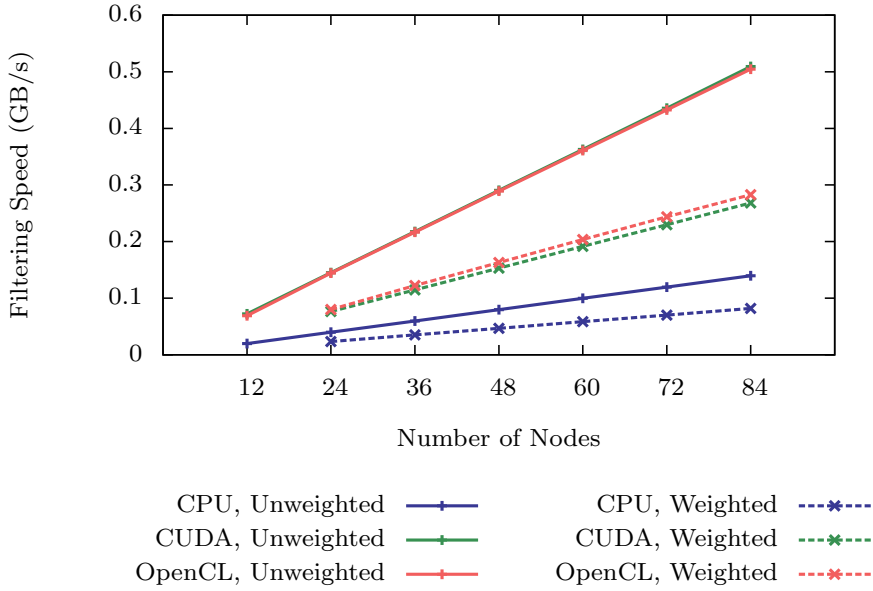}}
	\caption{{\bf Filter Processing Speed.} This plot shows the rate at which different implementations of the source finder filter the image data. This data only shows the time for the filtering stage of the source finder. It includes both the GPU kernel processing time and time spent transferring data between the host and the device.}
	\label{filter_speed}
\end{figure}

The filter speed can be separated out and analysed on its own to directly compare the different GPU implementations of the filtering process to each other, and the CPU implementation. This shows the difference in performance without the effects of the variable input times. The processing speed of the filter stage is shown in \figurename~\ref{filter_speed}. 

\section{Discussion}
There are a number of factors that contribute to the processing speed of the source finding program. Each step in the source finder has different performance characteristics and different resource requirements. After discussing the overall performance of the source finder, the most significant portions of the program will be considered.

\subsection{Overall Performance}
The real world performance of the source finder depends on all of the stages, and the hardware used to run the program. Measuring the overall performance of the program demonstrates the speed at which it will search a particular size of file. It is important to analyse the changes in performance as the program uses increasing numbers of nodes, in order to determine how effectively the program makes use of different amounts of resources.

We can see the performance of the program in \figurename~\ref{process_scaling}. This figure shows that the performance increases linearly with small numbers of nodes, but it starts to level off after 48 nodes. The drop off in performance is due to the increased amount of time spent reading the input data and increasing time spent transferring data between different processes, which will be discussed in Section~\ref{io_performance}.

\figurename~\ref{process_scaling} also demonstrates the differences in performance between the different implementations of the program, and the improvement gained from GPU acceleration. For unweighted filtering, OpenCL has a $1.8$ to $2.0$ times speedup and CUDA has a $1.8$ to $2.1$ times speedup, and in for weighted filtering OpenCL has a speedup of $1.7$ to $2.2$ times and CUDA $1.6$ to $2.2$ times.

These results shows that CUDA and OpenCL have very similar performance to one another, with the OpenCL implementation being slightly faster. This is to be expected as both CUDA and OpenCL implementations use the same optimisations and techniques. It should be noted that these results are true for the hardware and software configuration that is present of Fornax, and that the two implementations may perform differently to each other when the program is run on other hardware or software. Both of the GPU implementations are almost twice as fast as the CPU implementation, due to the GPU acceleration of the filtering algorithm.

\subsection{Filtering Performance}
The performance of the GPU filtering kernels can be affected by changing the parameters that control how they are mapped to the threads and processing cores of the acceleration device. The optimum values for these settings need to be found in order to get the best performance. This is done by performing a sweep of different values, and measuring the performance of each one. The first of the parameters is the block size used by the kernel.

The block size determines how many threads are working together using their shared memory to speed up data access. The speedup from this increases when using a filter with more elements along the frequency axis. Larger blocks will have a greater array of shared memory values to cache data, but by increasing the resource requirements of the block the total number of blocks that can be run at the same time is reduced, which lowers the thread occupancy. This means that the optimum block size is a balance between these two factors. The block size also affects the partitioning of the multiprocessor's resources among the different blocks that are running on that processor, which results in differing amounts of inefficiency due to not all of the resources being used.

The differences in performance can be seen in \figurename~\ref{group_size}. The variation up and down with each successive data point is due to how the evenly or unevenly the SM resources divide into the different blocks. As the block size increases, the partitioning of multiprocessor resources among blocks  that run on that multiprocessor becomes more coarsely-grained. This results in the drops in performance that appear for the larger block sizes. At the lowest block size the performance is decreased due to a limit to the number of blocks that can run on a multiprocessor, which combined with a very small block size limits the possible thread occupancy. After these effects, the amount of shared memory used by the blocks is not enough to significantly affect the performance of the kernels. There are several values that provide near-optimum performance for both kernels, including sizes of 128, 256, 320, and 512 threads, so choosing the exact optimum value is not crucial as long as it is from this set.

With an optimum block size determined, the effect of the total number of threads running on the GPU can be examined. The program can be configured to use different numbers of threads, on a scale from using a smaller number of threads doing a large amount of work each, to using a large number of threads doing a small amount of work each. Using a greater number of total threads increases the maximum amount of parallelism and increases the possible thread occupancy, but also incurs extra overhead when creating the threads. The results of changing the total number of threads used is shown in \figurename~\ref{ocl_total_threads} for the OpenCL implementation and \figurename~\ref{CUDA_total_threads} for the CUDA implementation. These plots show that the performance increases with the number of threads, although the effect diminishes. This figure also shows that it is not just the total number of threads that determines the performance, but also how many are along each axis. The performance can suffer when there are only a few threads being run along the X and Y axes, even if the total number of threads is high. From this we know that the largest number of threads gives the best performance, as long as the kernel processes a small enough portion of the data to avoid the watchdog timer, but that using a reasonably large number of threads is good enough.


The data shown here gives the information needed to get the best performance for the filtering kernels. It is important to note that the optimum parameters may change depending on a number of factors. These include the particular GPU or other accelerator device used for the program, and the version of any libraries and compilers used, particularly if new features are added. The optimum parameters may also change for different filter or data sizes, particularly as the program is optimised for filters that are largest along the Z direction. Using the optimum parameters for the hardware and software available on Fornax, the overall performance of the accelerated filtering algorithm can be measured.

The final speed of the filtering component, and how it varies with different numbers of nodes, can be isolated and examined independently, which is shown in \figurename~\ref{filter_speed}. This shows that the performance for each implementation increases linearly, across the entire range of nodes. This is as expected, because there is no communication between nodes in the filtering algorithm.

This figure also displays the difference in performance between different implementations. The CUDA and OpenCL implementations have similar performance to each other, with OpenCL generally performing slightly faster then CUDA, and both GPU implementations are significantly faster than the CPU-only implementations. When considering only the filters both implementations have a speedup of $3.6$ times the speed of the CPU implementation for the unweighted. For weighted filtering, OpenCL has a speedup of a factor of $3.4$ to $3.5$ relative to the CPU performance and CUDA has a speed up of $3.2$ to $3.3$ times the CPU performance. These results show that both GPU implementations significantly improve the performance of the filtering section of the source finder. The reason that the overall speed of the program shown in \figurename~\ref{process_scaling} does not follow the linear increases seen here is because other steps in the program do not experience a linear speedup with additional nodes. The most significant of these is the input step, which we will now discuss.

\subsection{IO Performance}
\label{io_performance}
The input time between runs can vary significantly, as shown in \figurename~\ref{input_time}. The main cause of different speeds for each run, and for each data point, is that the measurements were made at different times, when the storage system and the network were under different conditions. In some cases the input is significantly quicker, because it is already cached by the storage system. In other cases the input is particularly slow, because other jobs are using the storage system or the network. In particular, the large input times for $12$ and $24$ nodes are believed to be due to other jobs using the storage system at the time those tests were performed.

If the source finder is run in production on a shared supercomputing environment then those jobs will also have to share file system resources with other jobs on the system, and so they will also experience differences in input time. When running in such an environment it will be unlikely for the image data to already be in the system's cache, unless the source finder was being used as part of an image pipeline and directly used the data from the previous step. It is more likely that, on some occasions, the program will be running whilst other programs are using the storage system or network. Therefore the maximum input time from the data points shown here is the closest to being a representative value for the input time in a production environment.

There are some trends, however, that can be seen in the data in \figurename~\ref{input_time}. The input time increases at higher nodes, which is due to the image cube being broken up into a larger number of pieces, which results in a greater amount of halo data. This extra halo data needs to be communicated between the different MPI processes, increasing the transfer time. Additionally, the CPU input times tend to be slower than the GPU times for the same number of nodes. This is because the CPU implementation uses a greater number of MPI processes, which likewise increases the amount of halo data used by the system and so increases the amount of data that needs to be transferred.

The time required to read in data behaves differently than the computing time of other steps. \figurename~\ref{input_time} shows that, unlike the filtering step, the input time doesn't decrease with the number of processing nodes. This results in the input step taking a larger portion of the overall running time of the program as the number of nodes used, which can be seen in \figurename~\ref{input_time_portion}. The increase in time spent in the input stage of the program represents a waste of resources, because the program is still using the compute nodes but not improving the input time. Additionally, when the portion of the running time spent in the input step is large, the overall running time becomes more susceptible to slowdowns in reading the image data.


Because the input speed does not improve with additional processing nodes, and because the time spent in inter-node communication rises with the number of nodes, it is inefficient to run the source finder using a large number of nodes per job. It is better to use a large number of jobs that only use a few nodes each. Maximum efficiency is achieved by using the minimum number of nodes possible, which is set by the amount of memory required to hold all of the data needed by the program. It would also be beneficial to stagger the different jobs, so that they are not trying to read from the file system at the same time.

Whether the input step is a limiting factor on the speedup from increasing parallelisation is dependent on the speed that the image and weights files are read in. If the speed of the storage system were increased then the input step would not bottleneck the performance so much, and so the program would gain more from using additional computing nodes. Similarly, the use of a greater number of filters, filters that are larger in size, or other more computationally intensive algorithms will also increase the relative benefit from using additional compute nodes. This is because the input time does not change with the amount of processing work done to search the data, so as the amount of work increases the time taken to read in the data becomes less significant overall.

\subsection{Statistics Performance}
With the acceleration of the filtering step of the source finder, the other steps of the source finder take up a larger portion of processing time. These changes between implementations of PGSF can be seen by comparing Figures~\ref{filtering_time_portion}, \ref{statistics_time_portion}, and \ref{input_time_portion}, which respectively show the portion of time the program spends filtering image data, calculating statistics, and reading the data from disk.


In particular, the portion of processing time spent in the statistics and voxel selection steps of the GPU-accelerated source finder had increased significantly compared to the CPU-only implementations. This is because the while the filtering functions have been made faster using GPUs, the OpenMP acceleration of the statistics functions only improves their performance up to the same speed as the original, CPU-only implementation. Additionally, the statistics functions require communication between nodes, which increases processing overhead. Now that the statistics functions take almost as much time as the filtering functions, they are a candidate for future GPU acceleration efforts.

Overall, the GPU implementation of the filtering functions has significantly improved the performance of the source finder. Porting the filtering functions to GPU frameworks is more complicated than writing the same functions for CPUs, even when using CPU-based parallelism. There are more factors to deal with, including hardware layouts, transferring memory and technical issues. However, we were able to overcome these issues to produce a faster source finder.

\section{Conclusion}
This works shows that the use of GPU acceleration has significantly improved the speed of PGSF, for both weighted an unweighted analysis. The overall improvement in processing time is smaller compared to the improvement in the filtering step alone. This is because other steps in the source finder take the same amount of time, in particular the input step and the statistics calculations. These other steps are now larger contributors to the overall run time of the program.

Testing the program with different numbers of nodes shows that the program can make effective use of additional computing resources to reduce the processing time for a given data set, but it is more efficient to use fewer nodes per job. The inefficiencies are due to the time required to read in the image data, and to communicate between processes. These steps of the source finder cannot be improved through GPU acceleration or using additional computing nodes, so other techniques or hardware would be necessary to improve these stages.

There are some difficulties in porting source finding algorithms to GPUs. The code is more complex due to the highly parallel nature of the device, and there are more technical issues that must be dealt with such as separate memories and different hardware features. Additionally, if any other algorithms of a source finder are to be ported to GPU, a separate effort would be needed for each one. Advances in the hardware and software used in GPU computing may require that the algorithms used here be updated in order to achieved the best performance when using the new technology. But overall, GPUs provide a method of significantly improving the performance of source finders.

\subsection{Future Work}
The statistics functions could be ported to GPU, as they are now a more significant portion of the processing time of the program. The program could also be adapted to start processing part of a node's image data while, while the process is still reading the rest of the image data. This overlap would reduce the amount of time the program spends waiting for input, improving the overall speed. The current implementation could be modified to perform additional work per node but require less communication between nodes. It is possible, depending on the hardware used, that this could result in an overall decrease in processing time.

Another improvement is that the program could be expanded to include automatic configuration functionality to determine the best thread topology, and other performance parameters, what would save the user from having to determine these values themselves. Finally, if other algorithms and methods are added to SSoFF then it may be beneficial to also port those to GPU.

\section*{Acknowledgements}
This work was supported by iVEC through the use of advanced computing resources located at iVEC@Murdoch and iVEC@UWA.


\end{document}